Title: Chemical Complexity and Prevalence of Life in the Universe: A New Method for the Estimation of Key Terms of Drake Equation

Running title: Chemical Complexity and Life in the Universe

Author:         Lukasz Lamza
Affiliation:    Copernicus Center for Interdisciplinary Studies,
                Jagiellonian University
Address:        ul. Szczepanska 1/5, 31-011 Krakow, Poland
E-mail:         lukasz.lamza@uj.edu.pl

Abstract:
I describe a new method of estimating the prevalence of life in the Universe, based on the fact that more chemically complex environments are more rare. The paper makes three main claims: (1) There is a statistically significant (inverse) relationship between chemical complexity (quantified as the number of different types of molecules present in a given environment) and mass fraction for the successively smaller environments in the hierarchy of cosmic matter (extragalactic medium, interstellar clouds, dense cores, planetary systems, their icy fraction etc.) that is well described by a logarithmic law. (2) Minimal chemical complexity of life can be roughly defined, based on existing studies *in vitro* and *in silico*, both bottom-up (designing increasingly complex chemical systems) and top-down (simplifying minimal organisms). (3) Thus, one can estimate the fraction of the total mass density of the Universe that resides in reservoirs of chemical complexity estimated as being minimal for life. This is then translated, through simple statistical models of planetary systems, into the number of planets in a single Milky Way-sized galaxy that have, on their surface, reservoirs of "biogenic" chemical complexity. Two best models give the estimates of 1.6 (more complex minimal life) and $1.3 \times 10^4$ (slightly less complex minimal life) as the predicted upper bound for the number of instances of life per our Galaxy.

# 1. Introduction

There is considerable uncertainty concerning the possibility/prevalence of extraterrestrial life. For years, the standard reference point for discussing the "probability of life" has been the Drake equation. Its main advantage is, however, mostly its conceptual simplicity, and actually estimating the key terms of the equation, especially $n_e$ (the fraction of planets that might support life) and $f_l$ (the fraction of those planets that actually have life), is troublesome (see e.g. Vakoch, Dowd and Drake (2015), esp. the two reviews by Tirard and by Des Marais).

Here, I propose a new method of estimating the prevalence of life in the universe, based on a quantitative analysis of the qualitative observation that more complex chemical environments are more rare than simpler ones, and applying it to the problem of minimal chemical complexity of life. In short, a method is proposed for the calculation of the product $n_e \times f_l$ of Drake equation, and, with some additional assumptions, the upper bound for the number of planets with life per galaxy.

# 2. Methods and definitions

The basic idea behind the method advocated here is that more complex structures are less abundant than simple structures. In particular, the largest reservoirs of cosmic matter (i.e. the rare extragalactic medium) have relatively simple compositions, then increasingly smaller reservoirs (i.e. the warm interstellar medium, then cold molecular clouds, etc.) become more chemically complex. The trend continues within planetary systems, where gaseous objects (i.e. stars and gas giants) are both more massive and chemically more simple than solid planets etc.

## 2.1. Defining complexity

The general intuition ("more complex = less abundant") should now be made precise. This can be done in a number of ways, for instance:

1. In a natural mixture of chemical species ("reservoir") the complexity of a molecule (measured, for instance, by the number of atoms) is inversely proportional to its concentration.

2. In a reservoir there are fewer types of complex molecules than there are types of simple molecules.

3. In a reservoir that can be further divided into smaller reservoirs of different chemical composition, the larger reservoirs will have fewer chemical species than the smaller ones.

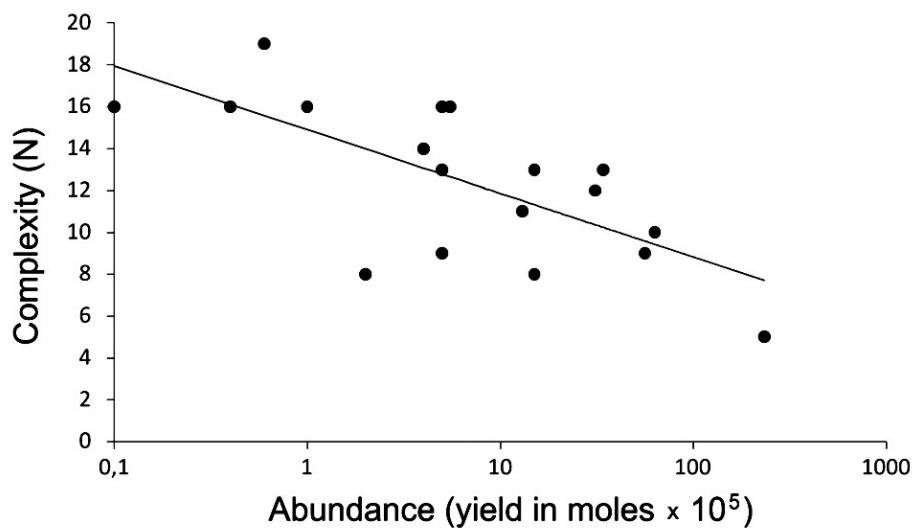

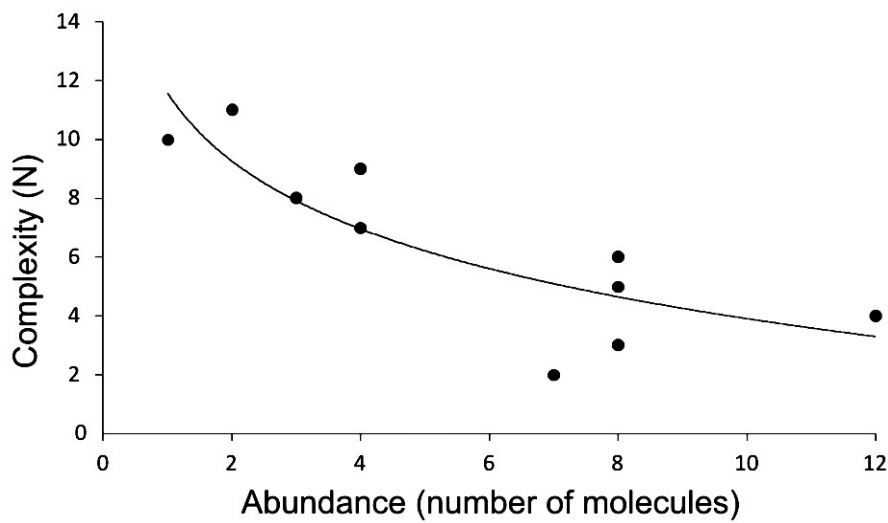

**FIG. 1 (a)** Abundance (defined here as yield in moles × $10^5$) versus complexity (here defined as the number of atoms in a given molecule, N) for the 18 chemical species described in the original Miller-Urey experiment (Miller and Urey, 1959), based on their Table 2 (p. 249). **(b)** Abundance (defined here as a number of different types of molecules) versus complexity (N) for the 57 chemical species discovered in the Taurus Molecular Cloud (Yamamoto, 2017). Both trend lines are logarithmic, but note that the horizontal axis is logarithmic in Fig.1a and linear in Fig. 1b.

  A good example that illustrates proposition 1 are the results of the original Miller-Urey experiment (Miller and Urey, 1959), where yields of compounds (their Table 2, p. 249) are clearly negatively correlated with the sizes of those compounds: larger molecules are much less abundant than small ones (see Fig. 1a). Similar results have been reported in actual astrophysical settings – see, for instance, the striking Fig. 3 in Sakai *et al.,* (2010, p. L52) where column densities of carbon-chain molecules in the Lupus molecular cloud are clearly inversely proportional to chain lengths, or Fig. 6.19 in Tielens (2005, p. 219), where sizes of aromatic molecules are inversely proportional to their abundances in interstellar space.

A simple example that illustrates proposition 2, in turn, is the number of chemical species of given size in the Taurus Molecular Cloud (Yamamoto, 2017). The full list of 57 molecules was sorted by size (number of atoms) and the number of different molecules of a given size was counted – see Fig. 1b.

In this paper, proposition 3 will be analyzed quantitatively. The whole Universe in the current cosmological epoch is divided into increasingly smaller sub-reservoirs of different chemical composition, based on standard distinctions in astrophysics. Then, for each such subdivision both its size (the fraction of total baryonic mass density of the Universe residing in that reservoir) and its chemical complexity (defined as the number of different types of molecules existing within it) will be estimated. There are tens of different methods of quantifying complexity (see e.g. Edmonds, 1999) and each has its own strengths and flaws. I argue, however, that for the purpose of this paper the one described above is sufficient: first and foremost, it is relatively easy to estimate, so can be a useful source of astrochemical and geochemical information. Also, an identical definition has been applied to the study of mineral complexity (Hazen *et al*, 2008). Furthermore, the sheer number of different types of molecules is argued to be only a "tracer" parameter that indicates a general process complexification of natural environments, which will be discussed in Section 6.

Some more clarifications are necessary at this point. A molecule will be counted as belonging to a different type only when the naked atomic "backbone" will be different. In other words, the method is "blind" to the state and number electrons. For instance, $CH_3^-$ and $CH_4$ will be counted as two separate molecules, but $CH_4$ and $CH_4^+$ – not. HCNO and HNCO are counted as two different molecules. Also, isotopologues will *not* be counted separately – $H_2$ and HD are still treated as a single type of molecule.

## 2.2. Defining life

The present paper makes essentially only a single assumption about life: that it is correlated with (but not necessarily causally related to, see Section 2.3 below) high chemical complexity. One might posit a slightly stronger definition: "life", to be subsumed under the reasoning presented in this paper, must realize its functions through chemical complexity. In other words, any hypothetical types of life forms that are *not* chemically complex are not counted in the present study. The controversies related to such a definition will be discussed in Section 6.

## 3. Cosmic chemical complexity

The key thesis of the article is that there is an inverse relationship between the size of cosmic chemical reservoirs and the number of different chemical species present in them. Although in cosmology, astrophysics and planetary science there are numerous "standard" subdivisions of cosmic matter, based on clear, physical principles, there are no "standard" estimates of neither their mass nor their chemical complexity, which had to be carefully assembled from numerous sources. This is described in details in Appendix A. Here, only the results are presented.

The Universe can be first divided into galaxies and the **extragalactic medium.** The galaxies, in turn, have two major components: the **stellar systems** (brown dwarfs, stars, and their remnants, and planetary systems that may encircle them) and the **interstellar medium** (ISM). ISM has a number of forms. Most of it is in a form of warm-to-hot plasma (which has no molecules), while parts of it condense and cools into overdensities, the clouds: first, the more irregular **diffuse clouds** that contain a fair number of simple molecules, then into better resolved, much cooler **molecular clouds** with considerable chemical complexity. Their inner parts, the **warm/hot cores**, contain even more molecules. The whole galactic medium is infused with microscopic **grains of galactic dust** which form a clearly separated phase characterized by peculiar chemical processes and species. When gas collapses gravitationally to form **stellar systems**, the central objects (stars) have only a small number of simple molecules in their atmospheres, while the **planetary system** is the locus of real chemical complexity. For the purpose of this article, a popular astrochemical division into the **gaseous**, the **icy** and the **rocky fractions** of the planetary system mass will be employed.

Tab. 1 gathers all the results from Appendix A. After regression analysis, the data are well fit by a logarithmic function, of the form $C_i = a \times \log(Y_i) + b$: see Fig. 2.

| reservoir | fraction of baryonic mass density, $Y_i$; $log(Y_i)$ in round brackets | chemical complexity, $C_i$ |
|---|---|---|
| 1. Extragalactic medium + intercloud ISM | 0.90 (-0.0458) | 0 |
| 2. Stars and stellar remnants | 0.077 (-1.11) | 7 |
| 3. ISM: diffuse clouds | 0.0069 (-2.16) | 30 |
| 4. ISM: (cold) molecular clouds | 0.004002 (-2.40) | 92 |
| 5. ISM: (warm/hot) dense cores | 0.000138 (-3.86) | 112 |
| 6. ISM: galactic dust | 0.00005 (-4.30) | 132 |
| 7. Planetary systems: the gaseous fraction | 0.0000963 (-4.02) | 39 |
| 8. Planetary systems: the icy fraction | 0.0000033 (-5.48) | 156 |
| 9. Planetary systems: the rocky fraction | 0.00000044 (-6.36) | 283 |

**Table 1**. Fraction of baryonic mass density ($Y_i$ and $log(Y_i)$) and chemical complexity ($C_i$) for the 9 reservoirs discussed in Section 3.

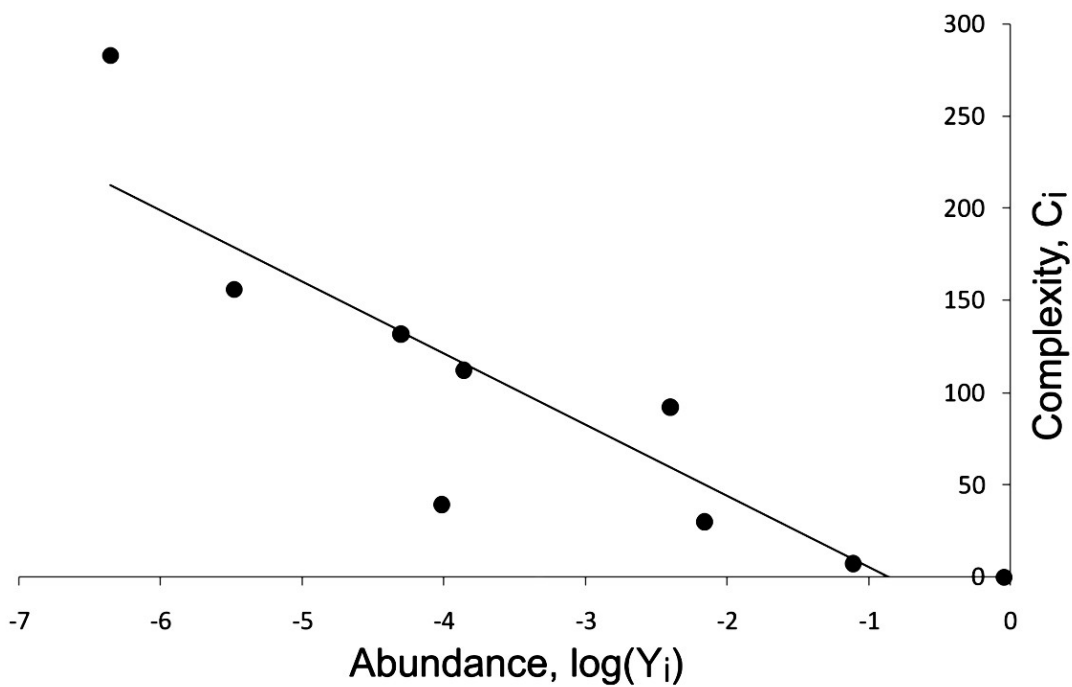

**FIG. 2** Abundance (defined as the fraction of baryonic mass density $Y_i$) versus chemical complexity ($C_i$) for the 9 reservoirs listed in Tab. 1: i=1...9. The best fit (black line) is: $C_i = -38.68 \times \log(Y_i) - 33.24$ ($R^2 = 0.74$, $p < 0.00185$).

**4. Minimal chemical complexity of life**

There is no general understanding of what constitutes a "minimal cell" and there is no generally accepted theory of abiogenesis. For the purposes of this study, however, a very specific estimate is needed – that of the number of different types of molecules in a simplest possible living system. There are two main roads towards a working estimate of that kind (see e.g. Luisi *et al.,* 1999): "bottom-up" and "top-down". The first begins with simple chemical systems which are known to be non-living, which are then further complexified – *in silico* or *in vitro* – to yield an object that would satisfy our definition of life. The second relies on known living cells which become progressively simplified – again, either conceptually or in actual laboratory practice – to produce "minimal cells".

**4.1. The bottom-up route**

There are numerous examples of theoretically conceived "minimal cells", starting with T. Gánti's classical chemoton, first published in 1971 (for a more modern discussion, see Van Segbroeck *et al.* (2005)), through E-CELL (Tomita *et al.*, 1999), often discussed as one of the first examples of a working whole-cell simulation *in silico*, to other models, both comprehensive (e.g. Forster and Church, 2006; Caschera and Noireaux, 2014) and modular (e.g. Castellanos *et al.,* 2004).

Such models usually have three main components that are of interest for students of cosmic chemistry: a minimal gene set, a corresponding set of reactions and a list of metabolites. The model can seem very simple, but a closer look always reveals that some of the chemical complexity has been swept under the rug. Let's take the example of E-CELL and the minimal cell by Forster and Church (2006) (henceforth: "FC cell"). The first has a small genome (based on *Mycoplasma genitalium*) consisting of 127 genes (of which 105 are protein-coding, 20 code for tRNAs and 2 for rRNAs) and concentrations of 84 "small molecules" are tracked. A simple calculation of the total number of different molecules gives 1 (DNA) + 105 (proteinaceous gene products) + 105 (their mRNAs) + 84 (small molecules) + 2 rRNAs = 297 molecules. The FC cell has 151 genes and – in contrast to E-CELL – is capable of reproduction and DNA maintenance, but has to be "fed" all basic metabolites, like sugars, lipids and amino acids, and cannot produce them. Altogether, a basic self-sufficient cell based on *M. genitalium* seems to need at least 150 genes[1], which is broadly consistent with the size of the minimal reaction set based on the metabolism of *M. genitalium* (Burgard *et al.*, 2001).

On closer look, the numbers are suspicious. First of all, such models tend to assume what Castellanos *et al.* (2004) refer to as "a maximally supportive culture environment". In practice that means that there are tens, if not hundreds, of molecules that are *known* to be necessary for a cell, that are assumed to be freely available to it and created in the environment. The study by Burgard *et al.* (2001) demonstrates it beautifully. Their minimal *M. genitalium* required just 122 reactions, but only in a "specially engineered growth medium", i.e. in one where, for instance, the TCA cycle is fully fueled by imported molecules, amino acids are imported etc. In a more realistic medium 224–229 reactions were necessary. From our perspective, however, the division of reactions and molecules into those occurring in the external environment, and those occurring within the cell, is completely irrelevant! The key point is that a certain number of molecules *does* have to be present for a very simple living cell to exist, and the question is how to estimate this number, not how to squeeze the smallest possible number into the cell and the largest possible number into the environment.

One last problem concerns the macromolecules. Their sheer existence inside our minimal cell is incompatible with the subject of the present study, which attempts to pinpoint a level of natural chemical complexification conducive for the formation of life, but *before* natural selection begins and life evolves to high levels of complexity. *All* proteins and nucleic acids included in the above calculations are products of natural selection. There are no conceivable scenarios in which an abiotic chemical environment spontaneously produces molecules of, say, translation initiation factor (one of the molecules that occurs in both E-CELL and the FC-cell), which is a protein composed of 72 amino acids that binds to mRNA and initiates its translation.

The problem is not trivial and, in fact, one might argue that it is in fact *the* problem of abiogenesis. Here, a radical step will be taken: *all macromolecules* will be left out of the calculation. What exactly that entails, we shall briefly discuss in the following section.

**4.2. The top-down route**

Let's begin with an estimate of the number of molecules actually present in simple cells. A standard textbook on molecular biology of the cell estimates that there are ca. 3821 different types of molecules in a "typical" bacterial cell (Alberts *et al.*, 2007) – see Tab. 2.

---

[1] Note that the "top-down" route seems to converge on a similar value. A classical study (Mushegian and Koonin 1996) gives 256 as the size of the "minimal gene set" derived from *M. genitalium*; a later analysis based on slightly different principles (Balasubramanian *et al.,* 2000) led to the estimate of 265 "essential genes".

| water | 1 |
| --- | --- |
| inorganic ions | 20 |
| sugars and precursors | 250 |
| amino acids and precursors | 100 |
| nucleotides and precursors | 100 |
| fatty acids and precursors | 50 |
| other small molecules | ~300 |
| macromolecules (proteins, nucleic acids, and polysaccharides) | ~3000 |
| TOTAL | 3821 |

**Table 2.** The number of types of molecules in a bacterial cell. After: Alberts *et al.* (2007, p. 55).

First, let's remove the inorganic ions from our calculation. Alberts *et al.,* don't provide a full list, but all of the examples they give on p. 63 are monoatomic ($Na^+$, $K^+$, $Mg^{2+}$...), so, according to the definitions established in Section 2.1., they should not be counted. As determined in Section 4.1 above, the macromolecules are to be left out as well. Proteins and nucleic acids have already been discussed. Polysaccharides have two main functions in bacterial cells: they form cell walls and are used for energy storage. We can leave them out and assume that simplest life doesn't need those two functions; also, polysaccharides form a minute percentage of all types of macromolecules.

What is left of "reasonably complex" molecules after macromolecules are out? It is worth noting that Alberts *et al.* define "small molecules" as ones having 30 carbon atoms or less (p. 55), which means that very simple polysaccharides, peptides and nucleic acids are actually subsumed under this definition. This may somewhat reassure the readers who are afraid that dropping "macromolecules" is too much of a simplification. In fact, it is now known that such "small macromolecules" can and do play important roles in living cells, and their formation in a prebiotic environment is drastically more probable than the formation of "full-length" proteins and nucleic acids that permeate modern cells. Oligonucleotides, for instance, play the key role in certain models of "RNA world" (see e.g. Orgel, 2004). The "small molecules", therefore are of key importance.

This leaves us with water and five sections: sugars, amino acids, nucleotides and fatty acids (plus their precursors) and "other small molecules". The total is now exactly 801 molecules including water, so this will be our first estimate for the minimal chemical complexity of life – it is the core chemical base of very simple, self-sufficient life forms that we know can exist.

How much can this estimate be reduced? It is reasonable, for instance, to assume that first forms of life on Earth didn't use all of the modern 22 proteinogenic amino acids. In fact, there are reports of *in vitro* simplification of the amino acid "alphabet", such as the study that, with only 13 amino acids, succeeds in creating "working" (foldable) proteins (Longo *et al.,* 2013). On the other hand, there is the known problem of amino acid chirality (see e.g. the book by Meierhenrich (2008)) which may mean that before L-amino acids were "selected" both forms were equally abundant. This goes for all molecules which are chiral and occur in modern cells only in specific isoforms. The numbers given by Alberts *et al.*, therefore, may be the result of not only a long period of the expansion of chemical complexity, but also of its pruning. Having said that, instead of discussing the 5 classes of small molecules and their possible simplifications – which would be not only tedious, but also introduce numerous new controversies, as it would in practice mean *inventing* various minimal forms of life – two simple estimates will be added. The first will be 75%, and the second 50% of the original value, representing a "somewhat simplified", and a "considerably simplified" set of, respectively, 601, and 401 molecules. A more detailed breakdown of possible savings is clearly beyond the scope of this paper.

Finally, there is one more important issue that we must address. All the above estimates refer to those molecules that are "part of life", be it within cells or necessarily present in their environment. However, in any realistic environment in the Universe, where life has evolved, we should expect the existence of molecules that are not part of the local biosphere. It is hardly conceivable that *exactly all* molecular species that existed on a surface of a given planet were incorporated into the first organisms. To use a specific example: the young Earth likely had at its surface all the types of molecules discussed in Section A.3.3 of the Appendix A, including polycyclic aromatic hydrocarbons, sulfur-containing heterocycles and tens of large carboxylic acids – all of which were *not* basis for first life, or, at the very least, did not *have to* be basis for first life. Of course, we shouldn't expect that all life in the Universe is biochemically equivalent to Earth life, but this only strengthens the argument. In the 'PAH world' scenario (Ehrenfreund *et al.*, 2006), for instance, the biochemical basis of early life are PAHs, and we should expect that in the hypothetical place in the Universe where this scenario actually unfolded, some of the basic molecules of Earth life, such as amino acids, would be passively present in the environment.

It is difficult to estimate the number of such "passive" molecules. Based on the expectation that life will most likely form in the most chemically complex phases of planetary systems, i.e. the icy and the rocky fraction (as will be later discussed in Section 5), which have chemical complexities of 156 and 283, respectively, a cautious value of half of their average will be assumed, i.e. 110. Therefore, to all three estimates of the minimal chemical complexity of life, at the final step the value 110 will be added as the "likely inert chemical background". This gives the following three numbers:

- 911: "high estimate"
- 711: "medium estimate"
- 511: "low estimate"

## 5. Results
### 5.1. Chemical complexity and reservoir size – a statistical analysis

The predictions of for fraction of baryonic mass, $Y_i$, based on the values of minimal chemical complexity of life discussed in Section 4, are listed in Tab. 3. This may be treated as the "raw" result of the present study.

|  | chemical complexity, $C_i$ | fraction of baryonic mass, $Y_i$ |
|---|---|---|
| (A) high estimate | 911 | $5.98 \times 10^{-20}$ |
| (B) medium estimate | 711 | $4.85 \times 10^{-16}$ |
| (C) low estimate | 511 | $3.93 \times 10^{-12}$ |

**Table 3.** Size of reservoirs (given as fraction of total baryonic mass of the Universe, $Y_i$) with the chemical complexity corresponding to three estimates for minimal chemical complexity of life discussed in Section 4. Predictions were made using the mathematical model from Fig. 2.

The unit in which the result is presented, $Y_i$, is not very practical. In order to convert it into something more useful, i.e. a predicted number of occurrences per galaxy, a number of further assumptions must be made, and a rigorous model of that kind would require a separate study on its own, and would still be debatable. Here, two simple models are introduced, intended as a proof-of-concept, rather than a definite solution.

The first (model A) is based on the assumption that life forms within planetary systems and is homogeneously distributed within them. The more refined model B is based on the narrower hypothesis that in such a case individual planets, or planetary-sized bodies ("planemos"), would be the actual seats of life formation.

### 5.2. Model A

For model A, let's assume that life forms within planetary systems, which, as we remember from Section 3, contain $Y_{PS}= 0.0001$ of total baryonic mass. This immediately gives column (i) of Tab. 4, i.e. the fraction of planetary system mass with the chemical complexities of $C_A$, $C_B$ and $C_C$. This can be translated into absolute mass (in kg), if we assume that our own Solar System is typical, with its mass of $M_{SS} \approx 2.76 \times 10^{27}$ kg, which gives column (ii) of Tab. 4, i.e. mass per planetary system.

|  | (i) fraction of planetary system mass (from Table 3); $Z_i = Y_i \times Y_{PS}$ | (ii) mass per planetary system; $M_i = Z_i \times M_{SS}$ |
|---|---|---|
| A. high estimate | $5.98 \times 10^{-16}$ | $1.65 \times 10^{12}$ kg |
| B. medium estimate | $4.85 \times 10^{-12}$ | $1.34 \times 10^{16}$ kg |
| C. low estimate | $3.93 \times 10^{-8}$ | $1.08 \times 10^{20}$ kg |

Table 4. Prediction for the occurrence of life in the Universe, according to model A.

Note that model A assumes that chemical complexity increases gradually and equally in all planetary systems. Therefore, the model also predicts that *each* planetary system will have, say, 33 kg of matter with the chemical complexity of 1500, and about 10 nanograms of matter with the chemical complexity of 2000.

### 5.3. Model B

The model A, however conceptually simple, is not compatible with what we know about the Universe and with the main mathematical model discussed here. Recall that the mass fractions $Y_i$ were always given for the *whole* reservoir of a given kind, only part of which would achieve the level of chemical complexity that is listed as the chemical complexity of the whole reservoir. For instance, interstellar dust grains are not *wholly* composed of the 132 molecules discussed in Section A.2.3; a large part of their volume and mass consists of mineral grains. The same goes, obviously, for stars (Section A.2.1), and all of the subdivisions of the planetary systems (Section 3.3).

In other words, the successive reservoirs are thought to be objects that *allow* and *facilitate* the emergence of a certain level of chemical complexity, never however fully transforming, in their whole mass, into objects of that complexity. For planetary-sized bodies, almost all chemical processes occur on or near the surfaces, and the bulk of the body partakes in chemical evolution indirectly – by supplying space, gravity field, heat, volcanic and tectonic activity etc. In that sense the whole planet is indispensable and is the "atom of cosmic chemical creativity", but does not attain in its whole mass the maximum chemical complexity that it allows.

Model B is based on the specific assumption that *planets* – in the sense of solid, "planetary-sized" bodies ("planemos" (Sarma *et al.,* 2008), i.e. those solid objects that are rounded by self-gravity), both rocky and icy – are the potential sites for the emergence of life, and that it is their total volume, and total mass, that should be thought as the basic unit of biogenesis in the Universe within the current model. More specifically, the term "planemo" includes bodies of diameter larger than ca. 450 km, which is the approximate limit, both theoretical (Weiss and Elkins-Tanton, 2013) and observational (i.e. actually true in the population of Solar System bodies), above which planetary gravitational field overcomes material strength and the object "flows" towards the spherical shape.

This is not just about shape: it is also at this approximate size (and for the same physical reasons) that planetary differentiation occurs, i.e. the body separates chemically into the core, mantle, crust and, if applicable, the hydrosphere and atmosphere. Note that the asteroid Vesta (average diameter 520 km) is rounded[2] and differentiated, Enceladus (504 km) is rounded and differentiated, Miranda (470 km) is *quite* rounded and *probably*, but not necessarily, differentiated and Proteus (420 km) is irregular and *likely* undifferentiated, i.e. it is a homogeneous body composed of only slightly reworked primitive Solar System material. Differentiation has *huge* consequences for chemical complexity, because it leads to an increase in concentration of numerous chemically important elements at the surface of the body.

The values given in column (ii) of Table 4 should now be translated into the number of individual planemos. Once again, we are forced to base our estimate on our Solar System, because there are no reliable counts of small bodies for other planetary systems. There are 24 solid rocky/icy planemos interior to the orbit of Neptune with a diameter larger than 450 km, and 95 trans-Neptunian objects of that size are currently known[3], which is probably not dramatically different from the actual number of such objects (Kenyon and Bromley, 2004). This gives 119 planemos with the total mass of $M_{tot} \approx 1.25 \times 10^{25}$ kg.

A rigorous calculation from $M_i$ (mass per planetary system; now interpreted as total mass of individual planemos with a given level of chemical complexity) to number of planemos would require a statistical model that relates planemo size (and/or other properties) to the level of chemical complexity, i.e. a continuation of the model derived in Section 3. This would require a separate analysis on its own, which is beyond the scope of the present study. Instead, two simple models are proposed.

First, if emergence of high chemical complexity is equally likely on all planemos, the median value of planemo mass (which is $M_{PL,med}=1.65 \times 10^{20}$ kg; i.e. between the mass of Enceladus and Pallas, the third largest asteroid) should allow us to calculate from $M_i$ to $B_i$, i.e. number of planemos with a given level of chemical complexity within a given number of planetary systems. For illustration purposes, the value will be normalized to $1\times10^{11}$, which is a reasonable estimate for the number of planetary systems in the Milky Way[4]. If, however, larger planemos are more likely to harbor large chemical complexity (after all, Earth is the heaviest of all 119 Solar System planemos, and, generally, larger ones do seem to be more chemically complex), some sort of statistical preference should be given to them. There is a number of ways to do it; probably the simplest is to employ, instead of median planemo mass, the arithmetic average, which is $1.05\times10^{23}$ kg (close to the mass of Jupiter's Callisto) – this introduces a desired statistical bias towards heavier objects.

---

[2] Or, to be more specific, *was* rounded before the impacts that excavated its southern hemisphere, forming the Rheasilvia and Veneneia craters.

[3] An up-to-date list of known trans-Neptunian objects is published by the Minor Planet Center (https://www.minorplanetcenter.net/iau/lists/t_tnos.html) and R. Johnston of the NASA Planetary Data System maintains a list with object masses aggregated from literature (http://www.johnstonsarchive.net/astro/tnoslist.html). Where object mass was not given, it was calculated from diameter, under the assumption of a spherical shape and the density of 1100 kg/m$^3$.

|  | (i) number of planemos with a given level of chemical complexity, per $10^{11}$ planetary systems, "flat model"; $B_i = 10^6 \times M_i / M_{PL,med}$ | (ii) number of planemos with a given level of chemical complexity, per $10^{11}$ planetary systems, "biased model"; $B_i = 10^6 \times M_i / M_{PL,ave}$ |
|---|---|---|
| A. high estimate | $1\times10^3$ | 1.6 |
| B. medium estimate | $8.1\times10^6$ | $1.27\times10^4$ |
| C. low estimate | $6.6\times10^{10}$ | $1.03\times10^8$ |

Table 5. Prediction for the number of planetary-sized objects ("planemos") that attain the given level of chemical complexity, per $10^{11}$ planetary systems. The "flat model" is based on the assumption that all planemos are equally likely to harbor higher chemical complexity. The "biased model" is based on the assumption that heavier planemos are more likely to harbor higher chemical complexity – see text.

**6. Discussion**

Let's now discuss critically the 3 main hypotheses of the paper:

1. *There is a statistically significant inverse relationship between reservoir size and chemical complexity in the Universe, i.e. higher levels of chemical complexity are consistently reached by a smaller fraction of cosmic matter.*

Based on data presented in Appendix A the hypothesis is demonstrated by statistical analysis in Section 3. However, one might argue that the data is flawed: both in terms of the choice of reservoirs, and the estimates for their size and chemical complexity. In defining the 9 reservoirs I have used standard distinctions employed in cosmology and astrophysics and I don't believe that there is reasonable ground for change. The second problem is more severe, because there are numerous specific sources of uncertainty discussed in Appendix A. Further studies will surely lead to the re-evaluation of the data in Table 1, however it is doubtful whether the general statistical tendency might be lost in such a process.

---

[4] The typical estimates for the number of stars in the Galaxy are in the range $1$–$4\times10^{11}$. The fraction of them having planetary systems is now usually estimated to be 0.25-1. Borucki *et al.* (2011) gave 0.34. Cassan *et al.* (2012) concluded that 17% of planets have Jupiter-like planets, 52% have Neptune-like planets, and "(p)lanets around stars in our Galaxy thus seem to be the rule rather than the exception".

A more specific problem is that of "trivial complexity". For instance, it is not inconceivable that there are in fact hundreds, if not thousands of subtle chemical variants of PAH molecules in interstellar space, and the emergence of a hundred new ones might not lead to the emergence of any new novel biochemical or biophysical processes. On the other hand, the formation of first pyrimidines and pyrines, even if it numerically entails only a dozen novel molecules, might lead to considerable increase in general complexity of the environment and be of great interest for astrobiologists. In other words, one might wonder whether the definition of "chemical complexity" employed in this paper really captures the increase in general complexity that has to accompany abiogenesis.

One way to mitigate this problem would be to define complexity not in terms of the number of types of molecules, but number of *classes* of molecules, using any of the numerous chemical classification systems. While this would introduce problems of its own (why couldn't life be highly specialized, i.e. be composed of numerous variants of just a single class of molecules?), I believe that it might make the sort of analysis presented in Section 3 much more robust in face of specific astrochemical controversies.

A second point is that, because of the possible existence of chemically rich environments that are clearly *not* alive, the values given in Table 5 should be understood as "upper bound for the number of planemos with life in a Milky Way-sized galaxy". Some of these planemos might harbor life, some of them – not. If, however, chemical complexity *necessarily* accompanies life, then all biospheres in the Galaxy are located on planemos whose number is included in Table 5. These values would then be true upper bounds.

2. *One can determine "minimal chemical complexity of life".*

First of all, it is important to emphasize that the reasoning is *not* that there exists a threshold of chemical complexity, above which life forms automatically, i.e. that high chemical complexity is the *cause* of abiogenesis. It would be probably safest to say that this is a (supposed) pure correlation, with zero causality implied, but in fact the reverse might be closest to the truth: that life *causes* high chemical complexity, even if it originally formed in an environment of low chemical complexity (such, for instance, would be the prediction of the mineral evolution theory (Cairns-Smith 1982)).

The proposition that life is *accompanied* by high chemical complexity is in line with a more general reasoning (AUTOCITATION) that formation of new types of molecules always signifies the emergence of novel natural structures and processes: chemical or otherwise. In other words, any place where 800 different types of molecules exist (with the possible reservation that "trivial complexity" should be accounted for), should be expected to exhibit highly complex dynamics of *some sort*. The most modest reinterpretation of the thresholds discussed in Section 4 would be that they represent not "minimal chemical complexity of life", but rather the level of chemical complexity that probably entails at least some processes typical for Earth life, such as formation of cells or metabolic control of chemical reactions. In other words, it might be argued that certain levels of chemical complexity must be accompanied by structures and processes so complex that they are within interest of astrobiology.

Of the three rough estimates discussed in Section 4, I believe that the largest one (the "high estimate") is the most realistic. Some of the reasons for that have been discussed in Section 4, but two arguments strike me as most relevant. First, the whole "macromolecular sector" has been removed, and for good logical reasons, but it still leaves open the question of how our putative "minimal life" is supposed to organize its dynamics. After all, the "gene-RNA-protein complex" in modern cells is responsible for essentially all of its functionality and "intelligence". Without saying anything particular about how metabolic control or heredity is supposed to be handled "generally" (i.e. in life as a cosmic phenomenon), it stands to reason that it should be accompanied by some chemical complexity. Second, all searches for "minimal life" on Earth, both *in vitro* and *in silico*, are invariably accompanied by sweeping at least some chemical complexity under the rug, rather than making it really obsolete. The "poster child" of biological simplicity, *M. genitalium*, is, after all, largely dependent on the metabolism of its host, and has actually *lost* the genes necessary for amino acid, nucleotide and fatty acid synthesis (Andrés *et al.*, 2011). Also, increasingly simple "virtual cells" are always simulated in increasingly chemically rich and generally more and more "controlled" environments. Such systems cannot be realistic models for first life, and if what we're looking for is not a biochemical-biophysical entity that is *barely* metabolizing under ultra-favorable conditions for a brief moment of time, but something that is robust, self-sufficient and can actually grow to become a biosphere, I believe that we must rather aim for the higher, not the lower estimates.

3. *Based on the results discussed above, one can calculate the abundance of life in the Universe and the number of planets that sustain life in a typical galaxy.*

At this point the weakest link of the reasoning are planetary science data that form basis of models A and B in Section 5. First of all, it is assumed that our Solar System is typical in the Universe. While 30 years ago this might have been laughable, now it's not unreasonable. Both the mass of our planetary system, and a list of its main components (gas giants, ice giants, rocky planets, debris disks, comets) are slowly seen to be quite typical for exoplanetary systems in the Solar neighborhood; some references supporting this claim have been cited in Section A.3. Overall, I don't see that assumption as a major source of error.

Model B, however, is based on a very specific assumption that individual "planemos" are the sites of abiogenesis in the Universe. Both scientific and science-fiction literature argues that this is not necessarily the case, and at the present state of debate in astrobiology it is prudent to agree. Planetary-based life, however, is definitely a possibility; after all, the only known case of life is of that very kind. Model B, therefore, openly restricts the reasoning to forms of life that emerge on the surface of planetary-size objects (whether it is planets, dwarf planets, or their satellites), and the "biased model" (column (ii) of Table 5) makes the specific assumption that it happens more likely on larger objects of that kind. All in all, I believe that the "biased model" is a much better approximation of reality, especially that is in line with the major argument developed in Section 3: larger planemos are considerably more rare than smaller ones, and should therefore harbor larger levels of chemical complexity on their surfaces.

Table 5 offers some interesting insights, especially that if the "biased" version of model B, and the highest estimate of "minimal chemical complexity of life" are to be trusted – which I consider to be the best combination of parameters – the result is a rather suggestive figure of 1.6. After taking into account all of the numerous uncertainties in the model, this study leads to the prediction that the spontaneous emergence of Earth-like life (i.e. based on chemical complexity and located on planetary surfaces) is not statistically impossible, but it should be expected to occur only on a handful of planets in a single Milky Way-sized galaxy.

**Acknowledgement**
The Author wishes to thank Lukasz Jach of the University of Silesia in Katowice for his help in performing the statistical analysis.

**Appendix A**

The starting point for the discussion of mass fractions will be the excellent, comprehensive review by Fukugita and Peebles (2004) – hereafter, FP2004 – a follow-up to an earlier article on the same subject (Fukugita *et al.*, 1998). Only in the case of planetary systems their estimation turned out to be considerably incorrect in the 14 years since the publication of the article. The chemical complexities of the successive reservoirs will be discussed based on standard review articles, if such exist. Preference will always be given to molecules actually observed in a given environment, as opposed to ones whose existence is deduced theoretically.

**A.1. The extragalactic medium**

Extended regions outside galaxies are poorly known, however it is now obvious that a considerable proportion of baryons is hidden there in the form of a rare, warm-to-hot medium. The "warm intergalactic plasma" (WIP) and the "intracluster plasma" (ICP) – the extremely hot ($T > 10^7$ K) medium between the galaxies within galaxy clusters – are usually treated separately in the literature, and so it was by FP2004 (see their Table 1). Although both of these divisions of the extragalactic medium are robustly non-molecular (Ryden and Pogge 2015, p. 239-241; Nicastro *et al.*, 2005), their size will be discussed in turn.

There is considerable uncertainty as to the exact proportion of cosmic baryons residing in the WIP. Dave *et al.*, (2001), in an analysis based on numerical simulations and preliminary observational constrains, propose a broad range 50-80%. Newer observations seem to be more consistent with the upper value, with Nicastro *et al.*, (2005) suggesting 88% (with a large uncertainty). FP2004 also give 88%, their result being basically all the other components subtracted from the total baryon density. To accommodate the estimates for the percentage of baryonic mass in both the ICP (see below) and in the galaxies (see Section 3.2), I have adopted the value of 85% here, and the chemical complexity of zero.

The mass of matter in the ICP is slightly better constrained than the mass of WIP, and seems to be equivalent to ca. 5% of the total baryon mass density (Dave *et al.*, 2001; Fukugita and Peebles 2014). The intracluster medium is typically thought to be a fully ionized plasma: "a nearly perfect gas of protons and electrons" (Cavaliere and Lapi, 2013), just as the diffuse intergalactic medium[5]. After adding their mass to the 85% of WIP, this gives 90% of cosmic mass in the extragalactic medium with zero chemical complexity.

**A.2. Galaxies and their components**

The percent of baryonic mass currently residing in galaxies is close to 10% (FP2004). Galaxies are made up of three main components:

---

[5] Recently, molecular lines of $H_2$ and CO have been described in extended regions around cluster galaxies, possibly reaching far into the ICP (Salomé *et al.*, 2011). They may be related to galactic outflows and/or the process of ram stripping (Jáchym *et al.*, 2014). The survival of those species in the conditions typical for ICP is dubious (molecules are likely to be destroyed by collisional dissociation and photoionisation), their continuous supply might however lead to a stable population of simple molecular species. As these observations are very preliminary, and seem to go against standard knowledge of ICP, they are disregarded in this study.

a) the interstellar medium (ISM);

b) the stars and their remnants (white dwarfs, neutron stars and black holes) and substellar objects (brown dwarfs);

c) "condensed objects" (i.e. dust, planets, asteroids etc.).

The masses, and chemical complexities, of those components, and their further subdivisions, are fortunately known much better than that of WIP and ICP. The values given by FP2004 are in good agreement with what can be found in newer literature (a reassessment of evidence by Fukugita (2011) reveals no significant discrepancies). According to their analysis, at the current cosmological epoch 23% of galactic mass resides in the ISM, 63% in stars and substellar objects (brown dwarfs), 14% in stellar remnants, and the remaining 0.2% is divided among the "planets and condensed objects" (together: 0.1%, described in detail in Section 3.3 below), and, not to be forgotten, the central supermassive black holes (SBHs) in the centers of galaxies (0.1%) that, because of their special nature, won't be considered further[6].

The term "condensed object", as used by FP2004, encompasses all objects held together by the electromagnetic forces, as opposed to gravity. The term "planets", on the other hand, encompasses those objects held together by gravity that were formed around stars (and either still orbit them or have escaped into the galactic space). Using rather simple arguments, the authors give 0.07% as the galactic mass fraction for condensed objects and 0.03% for planets. Their estimation was based on a very limited knowledge of exoplanets available at the time and it now seems to be slightly too low. Because these particular entries in the mass budget are of considerable importance for the present study, they will be discussed in more detail.

Draine *et al.*, (2007) in their study of dust emission in the 65 galaxies of the SINGS sample reach a range of dust-to-gas (D/G) values (0.003-0.01) that is consistent with that reported for the Milky Way and local dwarf galaxies (Galametz *et al.*, 2011) – which is typically in the range 0.001-0.01 (although there is a strong dependence of that parameter on galactic metallicity). Based on their data[7], 0.02-0.06% of total galactic (baryonic) mass resides in dust. While this result is broadly consistent with the value reported by FP2004 (0.07%), it is worth noting that the "condensed objects" of FP2004 also include all small bodies present *within* planetary systems, i.e. comets, asteroids and such, while the D/G ratios of Draine *et al.* are calculated based on emissions from the galactic ISM, and therefore include only objects floating in the interstellar space. Therefore, the fact that value in FP2004 is larger is fully expected.

In the present study a division much more reasonable in the context of cosmic chemical complexity will be used, i.e. *not* into the bodies held by electromagnetism and those held by gravity, but into (a) the galactic dust, and (b) all the condensed objects residing within planetary systems (whether held by electromagnetism or by gravity). In short, the value given by FP2004 for "condensed objects" must be decreased if we wish to calculate the mass fraction in galactic dust, because some of their "condensed objects" obviously reside in planetary systems, and a "concordance" value of 0.05% will be assumed for the percentage of galactic mass residing in interstellar dust.

---

[6] The chemical composition of SBHs, and any black holes, for that matter, has not been discussed in the literature, and common sense might suggest that there simply is no chemistry there. Common sense, however, is not necessarily the best guide in black hole astrophysics. For instance, SBHs may have densities comparable to the density of water (Celotti and Sciama, 1999), and there is no *a priori* reason why within their Schwarzschild radius there should not exist molecular species. Unfortunately, for obvious reasons, they remain unobservable, so SHBs, along with "normal" (stellar) black holes, will not be discussed further in this paper.

[7] The "naked" result by Draine *et al.,* (2007) was adjusted for the standard helium mass fraction (24% of total gas) – the "gas" in their D/G value refers to hydrogen atoms only – and the gas-to-total mass fraction of the galaxy (23%, discussed above).

Estimating the total mass of planetary systems within our Galaxy is not easy. As a first data point, let's note that 99.86% of our Solar System's mass is in the Sun, and the remaining 0.14% is in the planetary system (this figure is mostly determined by the mass of giant planets and is therefore very robust). If our planetary system is typical (which is not an unreasonable assumption – see below), then if 63% of galactic mass resides in stars and substellar objects, then ca. 0.09% is in planetary systems.

A different approach would be to start with protoplanetary disks, from which planetary systems form. Initially, they contain a sizeable fraction of the stellar mass, but gradually, in the timescale of 5-30 million years, they disperse, leaving behind the young planetary system. In a recent study of 176 young stars in our Galaxy's Taurus-Auriga region (Andrews *et al.*, 2013), a 'typical' disk-to-star mass fraction of 0.2-0.6% was established – a value slightly larger than the estimated mass fraction of our planetary system (0.14%), which is fully consistent with the fact that young planetary systems still lose some mass, by solar winds, photoevaporation and dynamical ejection (see e.g. Richling *et al.*, 2006). The lower value of 0.2%, translated into the fraction of galactic mass, gives 0.12%.

Actual observations of extrasolar planets (exoplanets) are becoming increasingly robust. It is now clear that most stars have planets, and that there is at least one planet per star in our Galaxy, and likely more (Cassan *et al.*, 2012). Note that, because of observational constrains, most extrasolar planet detections are of gas giants, which are the heaviest components of planetary systems (in our own planetary system they contain ca. 90% of the total mass). Estimating their mass distribution, however, is plagued by strong observational biases and there are currently no reported estimates with a high degree of statistical certainty.

Taking all the above into account, in the present work the following values will be adopted: 0.05% of galactic mass in interstellar dust grains and a "safe" value of 0.1 % in planetary systems.

In summary, the following values will be adopted:
- stars and remnants: 77% of galactic mass (7.7% of total baryonic mass of the Universe)
- ISM: 23% (2.3%)
- interstellar dust: 0.05% (0.005%)
- planetary systems: 0.1% (0.01%)

**A.2.1. Stars and stellar remnants**

Stars themselves can harbor a small suite of simple molecules in the cooler layers of their atmospheres (esp. in and just above photospheres) in their period of quiescent (main sequence) nucleosynthesis[8]. Based on a literature search, 7 molecules are consistently observed in the stellar photospheres (see e.g. Grevesse and Sauval, 1973; Asplund *et al.,* 2009): $C_2$, CH, CH, MgH, NH, OH and SiH. The situation is much less clear for stellar remnants, although simple molecules (notably $H_2$ and $C_2$) are known to be present in the atmospheres of white dwarfs (Koester, 2010; Hoard, 2012) and molecular hydrogen may exist in the atmospheres of neutron stars (Turbiner and López Vieyra, 2004). Because observations of the chemical composition of stellar remnants are of low quality, and because they are a minor contribution to the galactic mass budget when compared to ordinary stars, here they will all be subsumed under a single category, comprising 77% of galactic mass.

---

[8] It is worth noting that in any given moment a small subset of stars is in their last evolutionary stages, shedding outer envelopes (Asymptotic Giant Branch stars and planetary nebulae for lower-mass objects and Luminous Blue Variable-type objects and supernova shells for higher-mass stars) which is typically accompanied by spectacular chemical creativity. The inclusion of these objects, which would almost certainly be in line with the general thesis of the paper ("more complex-less abundant"), is one of the potential improvements of the present model.

**A.2.2. The interstellar medium**

The ISM is a much better-known environment than the extragalactic medium (see e.g. Ferrière, 2001; Yamamoto, 2017). From the chemical perspective, the main subdivision is into individual clouds, where chemical activity is concentrated, and the intercloud medium which is devoid of chemical activity.

52% of the ISM mass (i.e. 12% of the total galactic mass) is in the form of intercloud material which is robustly non-molecular (Tielens, 2005, p. 6-10), mostly due to high temperatures (T > 8000 K). This is confirmed by high-resolution, high-sensitivity maps of molecular lines, such as that of Andromeda galaxy (Nieten *et al.*, 2006). Because the rare outer layer of the galactic medium is continuous with the extragalactic medium and it is in impossible to unequivocally define the outer border of a galaxy, the intercloud ISM will here be added to the extragalactic medium to form the single, continuous, compositionally almost identical reservoir, encompassing 91.2% of the total baryonic mass density, and having zero chemical complexity, as defined in Section 2.1.

Traditionally, the so-called "molecular clouds" were thought to be the only places in the Galaxy that harbor chemical molecules. With time, it became clear that "traditional" molecular clouds (such as the Taurus Molecular Cloud) are embedded in less sharply defined "diffuse clouds" that show some chemical activity. This is especially well documented through the $^{12}$CO(1-0) line of carbon monoxide. Tielens (2005) gives 30% for the percentage of ISM mass budget in such diffuse clouds, and in two recent reviews of their chemistry (Tielens, 2001, p. 287–299; Yamamoto, 2017, Chapter 4), 30 molecules have been identified there.

Molecular clouds *sensu stricto* (also called "dark" or "cold" clouds) are well known for being very chemically rich (see e.g. Yamamoto 2017, Chapters 5-6). Different sources give different numbers, but they are usually close to that given in a recent review of molecular cloud chemistry, where 57 molecules were listed as robustly identified (Yamamoto 2017, p. 95-96)[9]. Molecular clouds contain ca. 18% of galactic ISM mass, according to Tielens (2005).

Some cold molecular clouds have even more dense, and slightly warmer "dense cores" (also called "hot"/"warm"/"lukewarm" cores, depending on how broadly they are defined), with temperatures significantly higher than the ambient molecular cloud, but usually still near or below 0°C, i.e. in the range of 15-300 K. In dense cores saturated molecules, such as dimethyl ether ($CH_3OCH_3$), many of them rich in nitrogen, are produced, plus some of the substances produced on the surfaces of interstellar grains (see Section A.2.3 below) can thermally dissociate from their surfaces and enter the ISM. At the same time, molecules typical for cold clouds, such as cyanopolyynes, still occur in dense cores (Chapman *et al.*, 2009). Herbst and van Dishoeck (2009, Table 1) specifically list 22 molecules present in dense cores, but *not* in the surrounding cold clouds. Comparison with the review of molecular cloud chemistry by Yamamoto (2017), that is used as the main reference in the present study, reveals that 2 of these molecules have since been identified in cold clouds (methyl formate and formamide), so the list shrinks to 20 additional molecules specific to dense cores, which gives the total of 77 molecules.

---

[9] The intensive search for complex organic molecules in molecular clouds have led to a number of highly controversial results, such as the putative existence of simple amino acids, incl. glycine and alanine. Also, the presence of certain species is often deduced rather than directly observed, and circumstellar molecules are often bunched with interstellar molecules (e.g. Herbst 2003), even though most of the former don't survive to the ISM. Such reports must be read with great care; here, only confirmed observations of molecular species in cold molecular clouds are counted.

To the best of my knowledge there are no published estimates of the mass percentage of molecular clouds belonging to dense cores. In our Milky Way, the ISM has the mass of ca. $5\times10^9$ $M_\circ$ (Draine 2009). Using the estimate by Tielens (2005) that 18% of ISM mass is in molecular clouds, this translates to $9\times10^8$ $M_\circ$ in molecular clouds. There are an estimated $10^4$ dense cores in our Galaxy (Lintott *et al.*, 2005) and the average mass of a core is ca. 3000 $M_\circ$ (Zinchenko *et al.*, 1998)[10], which gives 3.3% of molecular cloud mass in dense cores, i.e. 0,6% of total galactic ISM mass. Because Tielens (2005) did not discuss dense cores separately in his breakdown of galactic ISM, his value of 18% for molecular clouds must be now broken down into 17,4% in (cool) molecular clouds and 0,6% in (hot/warm) dense clouds, which are embedded inside them.

Note that polycyclic aromatic hydrocarbons (PAHs) were not included in the above literature sources. Because they are important components of the ISM, they will be discussed separately in Section A.2.4.

### A.2.3. Galactic dust

Interstellar grains, as they travel through the Galaxy, go through a cycle of chemical enrichment and depletion, while their ice mantles grow and shrink. In chemically rich environments, such as molecular clouds, they tend to absorb molecules from gas phase. In cool ISM the sticking coefficient – that determines how probable it is that a molecule impacting the grain surface will adhere to it – is close to one (Sandford and Allamandola, 1993), which means that all species present in gas phase can be expected to exist in and on grains. In short, the chemical complexity of interstellar grains is at least equal to the maximum complexity of the interstellar medium.

There is, however, a mechanism that increases the population of molecules on grains. If a chemical process occurs on a grain surface that produces a molecule which is *not* volatile, it will remain there, and therefore will be present only in interstellar dust, and not in the ISM. This forms the highly interesting fraction, variously termed organic residue (Allamandola and Hudgins, 2003) or refractory component (Greenberg *et al.*, 1995), embedded in interstellar grains, that later becomes incorporated into planetary systems, and is of great interest for astrobiology. (Of course, some molecules produced *in situ* on grains escape to space, with the notable case of $H_2$ and certain complex organic molecules (Biver *et al.*, 2014).)

Unfortunately, it is not easy to detect those molecules. As, by definition, they don't occur as free molecules in gas phase, they have to be detected through their imprint on the IR spectra of interstellar ices. Alternatively, one might simulate their formation in laboratory vacuum chambers, using analogues of interstellar ices (a typical example would be $H_2O:CH_3OH:NH_3:CO$ in proportions 100:50:1:1). This method yields interesting results and it is not unusual that amino acids (Caro *et al.*, 2002) or simple sugars (Meinert *et al.*, 2016) are produced in realistic interstellar ice analogues. Also worth noting are certain specific chemical "families", especially the POMs (polyoxymethylenes, i.e. molecules formed through the polymerization of formaldehyde, containing the –($CH_2O$)– monomer) and a family of ringed polymers called carbon suboxyde polymers (Ballauf *et al.*, 2004) that might form in interstellar ices. High resolution mass spectra suggest that tens of different types of molecules can form in such experiments (Allamandola and Hudgins 2003, p. 299), but – because no certain detections have been claimed so far – a conservative value of 20 will be assumed here: probably the least constrained of all values discussed in this article.

---

[10] Average dense core mass, based on their Table 4 (p. 345), which is the data set for "category I" cores, i.e. those whose parameters are best constrained.

**A.2.4. Polyciclic aromatic hydrocarbons (PAHs)**

Polycyclic aromatic hydrocarbons (PAHs) deserve their own section, because of both their large potential chemical complexity and the extremely high level of uncertainty surrounding the subject. PAHs are molecules having the form of fused benzene rings, or their derivatives, and are mostly formed in hot stellar outflows (Tielens *et al.*, 2000) (although formation *in situ* in ISM is sometimes discussed (Parker *et al.*, 2012)). They are problematic in that: (a) it is now clear that they are an important part of the galactic carbon budget (Allamandola and Hudgins 2003); (b) tens to hundreds of different types of PAHs and their derivatives are known – spectral databases, created explicitly to help astronomers in detecting interstellar PAHs, list anywhere from 40 (Mallocci *et al.*, 2007) to 583 (Bauschlicher *et al.*, 2010) different molecules (the latter list also includes oxygen and nitrogen-containing derivatives); (c) it is very difficult to translate from observational data to individual molecules (Tielens 2005, p. 316), as usually only very non-specific signatures are available (such as infrared signatures of the aromatic C-H stretching mode or various ring bending modes (Tielens *et al.*, 2000; Tielens, 2005, Table 6.5 on p. 215)).

On the other hand, there are good reasons to be skeptical about the actual number of different PAHs present in ISM and on interstellar grains. First, it is not certain whether individual PAHs can actually survive for a long time in the ISM[11], with some authors forcefully arguing that it's in fact impossible (Micelotta *et al.*, 2010), and even their stability within interstellar ices may be dubious (Gudipati and Young, 2012). It may well be that through dissociation, fusion with neighboring PAHs, "amorphization" of benzene subunits and other processes, a large portion of what is suspected to reside in PAHs is actually in the form of a complex, irregular, continuous substance, variously called soot, disordered graphite or glassy/amorphous carbon (Herbst and Van Dishoeck, 2009). It is worth noting that even silicate grains in the interstellar medium become quickly "amorphized" (Carrez *et al.*, 2002) – some estimates say that >99% of all silicates in ISM are amorphous (Kemper *et al.*, 2004). In short, the ISM is not a hospitable environment for regular, highly organized chemical structures.

It is usually suggested that PAHs with ca. 40-70 carbon atoms are responsible for most of the interstellar infrared emissions (Tielens *et al.*, 2000) which might suggest that "monster" PAHs, the size of hexabenzocoronene ($C_{42}H_{18}$) or circumcoronene ($C_{54}H_{18}$) exist in ISM. In reality, however, it is not clear whether the IR signal comes from individual molecules of that size, or rather from weakly bound clusters of smaller PAHs and/or their derivatives (Tielens 2005, p. 217). If that second case was true, the largest individual PAHs might be, for instance, the size of coronene ($C_{24}H_{12}$).

The scope of the chemical modification of PAHs is also controversial. Theoretical calculations suggest that PAHs can exist in a number of derivative forms and that hydrogen atoms may be substituted by certain side groups, such as the methyl, hydroxyl or aldehydic group. In fact, careful modeling of interstellar IR bands leads to the conclusion that >95% of all side groups in PAHs are actually hydrogen atoms, i.e. the "basic" form (Tielens 2008).

---

[11] When PAHs are robustly detected in ISM, it is usually in the gaseous shells surrounding evolved stars, i.e. place of their formation – see e.g. Mulas *et al.* (2006). PAHs, or their chemical modifications with aliphatic side-chains, may be responsible for the Diffuse Interstellar Bands (DIBs), which are notoriously difficult to identify (Cordiner, 2011). The future identification of DIBs may considerably change our understanding of the chemical complexity of ISM.

In summary, it remains to be established how many types of PAH molecules are actually present in astronomical environments. However, because their detection in interstellar ice, cometary material (Sandford *et al.*, 2006) and in meteorites (Elsila *et al.*, 2005) seems to be secure, it is necessary to include them in the calculations. A cautious approach will be applied, consistent with the general methodology of this study, to follow empirical detections, if such are available, rather than theoretical calculations. 7 PAHs have been actually produced through the irradiation of $CH_4$ ice by Kaiser and Roessler (1997), and – to accommodate the intermediate steps in their synthesis (Bauschlicher and Ricca 2000) – a maximum of 18 additional molecular species must be present, likely less[12]. After allowing for some side group substitutions, a value of 30 different PAHs will be assumed here as representing a reasonable "safe" estimate based on actual observational data rather on what's theoretically possible.

It is worth noting that this value is fully consistent with the number of PAHs found in the organic residue of carbonaceous chondrites (see Section A.3.3. below): the review by Sephton (2002) cites the detections of 23-34 different PAHs in 3 unrelated studies employing different methods on different rock samples. Because there are no data solidly indicating the difference in composition between the PAHs in the interstellar environment and inside planetary systems – and the latter are genetically related to the former (Charnley and Rodgers 2008) – it will be assumed here that the value of 30 is valid for all the 3 environments where data about specific PAH species are not available, i.e. cold molecular clouds and their hot dense cores (but not diffuse clouds which have no PAHs – see: Tielens (2006), p. 323) – and the galactic dust. Note that PAHs are also present in comets and the icy bodies of the outer Solar System (see Section A.3.2), in atmospheres of gas giants (Sagan *et al.*, 1993; see Section A.3.1), and in rock samples from asteroids and rocky planets (see Section A.3.3).

**A.3. Planetary systems and their components**

Planetary systems are clearly of key importance when it comes to cosmic chemical complexity. Unfortunately, only a single planetary system is well known which makes estimating the mass budget of the sub-sections of planetary systems highly speculative. On the other hand, it is becoming increasingly clear that many exoplanetary systems are similar to our own when it comes to basic structural elements. We know that both gas giants, icy giants and rocky planets can co-occur in exoplanetary systems (Cassan *et al.,* 2012), that debris disks analogous to the asteroid belt and the Kuiper belt can form (e.g. Chen and Jura, 2001) and analogues of Oort cloud have been tentatively detected (e.g. Kiefer *et al.*, 2014).

If theories of planet formation and cosmochemistry are to be believed, certain basic properties of our Solar System should be thought of as generic (Klahr and Brandner, 2006; Kennedy and Kenyon, 2008), and it is indeed the most parsimonious assumption. For the purpose of the present paper I have decided to distinguish the three above-mentioned "fractions" of planetary system: the gaseous, the icy and the rocky/metallic.

---

[12] The reasoning is as follows. Out of three main methods for PAH growth discussed by Bauschlicher and Ricca (2002) the "cationic route" is selected by the authors as most relevant for the conditions in ISM (as opposed, e.g., to terrestrial flames). This route predicts three intermediate species between two successive PAHs, however this is only for the case of "linear" growth, where the new ring shares only two carbon atoms with the existing structure. In actual PAHs produced by Kaiser and Roessler (1997) this is not always the case, and less intermediates will be necessary to produce, for instance, the last ring of perylene (see their Fig. 2 on p. 147). If ring fusion is a possible route to PAHs, the number of intermediates is even smaller. Hence, the 18 intermediates between the 7 PAHs of Kaiser and Roessler (1997) – 3 intermediates per 6 steps – is just the upper bound.

An important remark is at place. Because most bodies are composed of a mixture of those fractions, here a given body will be classified as gaseous, icy or rocky/metallic according to which of those three type of material dominates at the surface layers of the body. This seems to be the only reasonable decision when discussing chemical complexity. The case of giant planets should clarify the issue. Jupiter and Saturn are classified as gas giants, and Uranus and Neptune as ice giants, based on their total chemical composition. There is, however, hardly any "ice" on Uranus and Neptune. The thick outer layers of both of these planets are gaseous, chemically almost identical to those of Jupiter and Saturn, and any "ices" (here: certain simple compounds of oxygen, carbon and nitrogen) are squeezed and heated under thick layers of atmosphere, and are in a superheated, conductive fluid state that has nothing to do with the behavior of "ice", both physically and chemically. In fact, under temperatures and pressures assumed to exist in the "icy" mantles of Uranus and Neptune there is precious little chemistry – after all, it is a supercritical liquid with temperatures no lower than 2000 K (Cavazzoni *et al.*, 1999). Consequently, chemical phenomena known to occur within those planets are much more comparable to what's happening in the atmospheres of giant planets than on the surfaces of truly "icy" bodies such as Pluto or Rhea. Furthermore, it is important to mention that the "chemical complexity" of a given reservoir is not typical of the *whole* reservoir: a point that will become crucial in Section 5.3.

Consequently, in our Solar System the gaseous fraction would include Jupiter and Saturn, *and* Uranus and Neptune; the icy fraction: the Oort cloud, and all minor bodies with icy surfaces (most outer Solar System satellites, dwarf planets and other trans-Neptunian objects), while the rocky/metallic fraction: the four rocky inner planets, the asteroid belt and all rocky satellites and other minor bodies of the outer Solar System. The masses of all abovementioned objects or groups of objects are very well constrained, save for the trans-Neptunian objects. Here, I shall assume the estimate by Gladman *et al.* (2001) that the total mass of the Kuiper belt (understood broadly to also include scattered disk objects and centaurs) is 0.1 $M_E$ (Earth's mass). The mass of Oort cloud is even worse constrained[13]. Here, a "conservative" total mass of Oort cloud of 15 $M_E$ will be assumed.

After summing up all the known components of our planetary system (i.e. Solar System minus the Sun), we reach a total mass of ca. 462 $M_E$, of which 445 $M_E$ (96.3 %) reside in the four "gaseous" giants, 15 $M_E$ (3.3 %) in all icy bodies, and only 2 $M_E$ (0.44 %) in all rocky bodies.

**A.A.1. The gaseous fraction**

The gaseous atmospheres of giant planets are fairly well-known in terms of their chemical composition. In a review and comparison of atmospheric chemistry of Jupiter and Saturn (Atreya *et al.*, 2003) most molecules are listed to occur in both planets; the full list includes 21 species (their Table 1). In order to account for the benzene and PAHs observed in Jupiter's polar regions (and likely Saturn's (West *et al.*, 2009)) ca. 15 further species need to be included (Wong *et al.*, 2000). The chemistry of the gaseous envelope of Uranus and Neptune is highly similar to the ones of Jupiter and Saturn, and a comparison of a list of molecular species found in the atmospheres of these two ice giants (Lodders and Fegley, 2015, Table 7.3, p. 408) with the list in Atreya *et al.* adds only 3 (nitrogen-containing) species. This gives the total of 39 molecules.

**A.A.2. The icy fraction**

---

[13] In the comprehensive review by Stern (2003) a range of 1-50 $M_E$ is given. Dones *et al.* (2004) cite values of 3.3-7 $M_E$ for the mass of the outer Oort cloud (a > 20000 AU), which they find better constrained. Kaib and Queen (2009) give 6.7 $M_E$ for the mass of the inner Oort cloud (3000 AU < a < 20000 AU), while noting that a significant reservoir of comets may exist in ever "more inner" parts of the Oort cloud (a < 3000 AU).

By far the largest reservoir of the icy fraction of our planetary system is the Oort cloud, but the chemical evolution of larger objects, such as those inhabiting the Kuiper belt, is qualitatively similar, and they are often discussed together (e.g. Prialnik *et al.,* 2008). Comets are evolved, but direct descendants of interstellar grains – in this sense they are "the most pristine remnants of the formation of the solar system" (Biver *et al.,* 2014). The actual cometary nuclei of the Oort cloud are notoriously difficult to observe, however as soon as one enters the inner Solar System and starts shedding its outer layers, a bewildering richness of molecular species becomes apparent. Despois *et al.* (2005) list 48 confirmed molecular species found in comet tails. There is also, however, an organic-rich residue, composed of what is often referred to as "CHON" particles (sometimes extended to "CHON-PS"), that amounts to, by one estimate, 23% of cometary nuclei by mass (Greenberg, 1998). Chemically, it is thought to be broadly similar to the organic residue in the ice mantles of interstellar grains, however comets go through a much more complex processing, both by radiation, and by heat (Prialnik *et al.,* 2008). It is likely that at least *some* cometary nuclei, because of radioactive heating, may for a time have pockets of liquid water (Wooden 2008) which would, obviously, greatly increase their chemical complexity, and is something clearly impossible in interstellar space.

Finally, there are planetary-sized bodies, sometimes called "planemos" (Sarma *et al.,* 2008), i.e. those large enough to have been rounded by gravity. At the moment, ca. 120 of these are known to exist in our Solar System, most of which have considerable ice crusts and icy surfaces, and at least a few likely have subsurface oceans of liquid water. Their chemical complexity is only beginning to be known, however it is worth noting that volatiles shed by these bodies, such as those that make up the plumes of Enceladus (Matson *et al.,* 2007), are compositionally broadly similar to cometary tails. There are no comprehensive reviews of the chemistry of icy satellites, so the value of 48 given by Despois *et al.,* (2005) will be used for the chemical complexity of the volatile fraction of icy bodies.

Finally, many icy bodies contain rich, usually reddish-to-brown organic residues on their surfaces, often discussed under the umbrella term "tholins". Tholins have been long known to exist in the ices of numerous outer Solar System bodies (Grundy *et al.*, 2002; Cruikshank *et al.,* 2005), icy particles of Saturn's rings (Cuzzi and Estrada, 1998), and recently have been observed by *New Horizons* on Pluto and Charon (Grundy *et al.,* 2016). The exact chemical nature of tholins is not known. Studies suggest the occurrence of numerous complex molecules with very high N/C proportion, likely containing N-C heteroaromatics, with masses up to 500 daltons (Quirico *et al.,* 2008), which is especially interesting from the perspective of astrobiology. In an earlier study (McDonald *et al.*, 1994), a simulated Triton tholin contained 15 amino acids and a broad, unresolved population of molecules containing 10-50 atoms of C+N (combined). In a more detailed study of Titan tholin analogue (Somogyi *et al.,* 2005) 108 different molecules were found (their Table 2). This value seems reasonable as a first approximation. Adding to the 48 confirmed detections in gaseous phase, this gives the total of 156 molecules in the icy fraction of planetary systems.

**A.A.3. The rocky fraction**

At this point a strong potential source of bias must be taken into account: it is clear that the present summary of cosmic chemical complexity is written from a place that is "selected" to be highly chemically complex. While it has been argued in Section A.3. that Solar System's general structure, and the chemical richness of its gaseous and icy bodies *may* be representative of what's typically happening in the Universe, Earth cannot be assumed to be typical. One solution to the problem is to account for the total chemical complexity of the rocky/metallic fraction *minus* Earth, and this will be in fact pursued here.

Even after excluding Earth the rocky fraction of the Solar System has quite complex chemistry. Studies of meteorites, notably carbonaceous chondrites, reveal a stunning diversity of chemical species, and in a recent reappraisal of one of the most famous meteorites of that kind, Murchison, the authors argue that "tens of thousands" of different molecules may be theorized to exist between its mineral grains (Schmitt-Koplin *et al.,* 2010). For decades, molecules such as hydrocarbons and amino acids (Kvenvolden *et al.,* 1970), and even triazines and purines (Hayatsu *et al.,* 1975) have been detected in Murchison and other meteorites. Also, some Martian meteorites may contain complex chemical species, including aliphatic and aromatic hydrocarbons, phenol and benzonitrile (Sephton *et al.,* 2002). An interesting development has been the recent discovery of "active asteroids", also – probably more inspiringly – called "main-belt comets" (Hsieh and Jewitt, 2006): asteroids that present comet-like tails of dust and volatiles. Jewitt *et al.* (2015) list 18 such bodies and it is clear that their study will bring considerable advance in our understanding of the chemical composition of Solar System's rocky bodies.

Here, molecules found in carbonaceous chondrites will be treated as a proxy for "typical" chemical complexity available for rocky bodies in the Solar System and, in fact, planetary systems as such[14]. Carbonaceous chondrites are typically seen as a pristine reservoir, reasonably well representing the rocky material that filled the inner parts of the early Solar System (Hutchinson, 2004, p. 97). In 2002 an impressive review of organic compounds (that also includes simple inorganic compounds) found in carbonaceous meteorites was published (Sephton 2002). Although the author does not give a single figure for the total number of different molecules, his Table 2, which lists 25 classes of organic compounds, combined with highly detailed discussion of those classes and a rich list of references, allows for a quite specific estimation that 283 molecules have been detected in carbonaceous meteorites[15], and this value will be assumed in the present study.

---

[14] Although mineralogical and chemical compositions of exoplanets are still largely unknown, there are interesting reports that certain extrasolar asteroids may have elemental compositions similar to CV chondrites (Jura *et al.*, 2012). Overall, processes responsible for the formation of chondrites are expected to be universal in character (see: Hewins *et al.,* 1996, esp. the chapter by L. Hartmann, pp. 13-20).

[15] Because this estimate is so important for the present study, a detailed breakdown might be handy. All the reference numbers and page numbers in this footnote refer to Sephton (2002). The number of different molecules for all the classes in Table 2 that are not monotypic (such as "Carbon monoxide" which forms a separate class) are as follows: **0** aliphatic hydrocarbons (a very strong argument for terrestrial contamination is presented on pp. 297-299; this controversial issue is also discussed in: Sephton *et al.,* 2001), **34** aromatic hydrocarbons (the top value of the range 23-34 for sources cites on p. 295), **26** monocarboxylic acids (ref. 67), **40** dicarboxylic acids (a safe estimate after ref. 78 which says "more than 40")), **56** hydroxycarboxylic acids (Tab. 2 from ref. 78), **17** amino acids (the top value of the range 7-17 for sources cited on p. 299-300, save for the single value "more than 70" which was given in reference 120; in this paper, however, no specific list of amino acids is presented, and the value itself is not supported by further references), **4** alcohols (p. 304), **5** aldehydes (p. 304), **5** ketones (p. 304), **19** sugar-related compounds (Fig 1A and 1B of ref. 34), **10** amines and **3** amides (refs 35 and 141), **11** N-heterocycles (4 purines and 2 pyrimidines mentioned on pp. 304-5, plus 5 "definite identifications" of (iso)quinolines+pyridines in ref. 38), **7** pyridinecarboxylic acids (ref. 39), **4** dicarboximides (ref. 39), **10** benzothiophenes (ref. 42), **7** sulfonic acids and **5** phosphonic acids, plus the sulfate ion and the phosphate ion (ref. 44, Table 1) and **12** S-heterocycles (not listed in Sephton's Table 2, but discussed in the article anyway (ref. 56, Table 3)). After adding all the 5 "monotypic classes" ($CO_2$, CO, $CH_4$, $NH_3$, $CO(NH_2)_2$) this gives 282 molecules. Water wasn't listed separately by Sephton, but is obviously present, which gives the total of 283 molecules.